\documentclass[twocolumn,showpacs,preprintnumbers,amsmath,amssymb,nofootinbib
,superscriptaddress]{revtex4}
\usepackage{hyperref}
\usepackage{graphicx}
\usepackage{color}

\newcommand{\be}{\begin{equation}}  
\newcommand{\ee}{\end{equation}}  
\newcommand{\bear}{\begin{eqnarray}}  
\newcommand{\eear}{\end{eqnarray}}  
\newcommand{\ba}{\begin{array}}  
\newcommand{\ea}{\end{array}}

\newcommand{\sparen}[1]{\left[ #1 \right]}
\newcommand{\bparen}[1]{\left\{ #1 \right\}}

\newcommand{\vev}[1]{\left< #1 \right>}

\newcommand{\parenfrac}[2]{\left( \frac{#1}{#2} \right)}

\newskip\humongous \humongous=0pt plus 1000pt minus 1000pt

\newif\ifdtup

\def\oldreffmt#1{\rlap{[#1]} \hbox to 2\parindent{}}

\def\figfmt#1{\rlap{Figure {#1}} \hbox to 1in{}}  
  
\def\ie{\hbox{\it i.e.}{}}	  
\def\eg{\hbox{\it e.g.}{}}

\def\vev#1{\left\langle #1\right\rangle}

\def\beq{\begin{equation}}  
\def\eeq{\end{equation}}  
\def\bea{\begin{eqnarray}}  
\def\eea{\end{eqnarray}}

\def\bq{\begin{quote}}  
\def\eq{\end{quote}}

\relax  

\newdimen\tdim  
\tdim=\unitlength  
\def\bar{\overline}

\begin{document}

\title{An ultra-weak sector, the strong CP problem and the pseudo-Goldstone dilaton}

\author{Kyle Allison}
\email{k.allison1@physics.ox.ac.uk}
\affiliation{Department of Theoretical Physics\\
University of Oxford, 1 Keble Road\\
Oxford OX1 3NP\\$ $}

\author{\\Christopher T. Hill}
\email{hill@fnal.gov}
\affiliation{Fermi National Accelerator Laboratory\\
P.O. Box 500, Batavia, Illinois 60510, USA\\$ $}

\author{\\Graham G. Ross}
\email{g.ross1@physics.ox.ac.uk}
\affiliation{Department of Theoretical Physics\\
University of Oxford, 1 Keble Road\\
Oxford OX1 3NP\\$ $}

\date{\today}

\begin{abstract}
In the context of a Coleman-Weinberg mechanism 
for the Higgs boson mass, we address the strong CP problem.
We show that a DFSZ-like 
invisible axion model with a gauge-singlet complex scalar field $S$, whose
couplings to the Standard Model are naturally ultra-weak, can solve the 
strong CP problem and simultaneously generate acceptable electroweak symmetry 
breaking. The ultra-weak couplings of the singlet $S$ are associated with 
underlying approximate shift symmetries that act as custodial symmetries 
and maintain technical naturalness.  The model also contains a very light 
pseudo-Goldstone dilaton that is consistent with cosmological Polonyi bounds,
and the axion can be the dark matter of the universe. 
We further outline how a SUSY version of this model, which may be required 
in the context of Grand Unification, can avoid introducing 
a hierarchy problem.

\end{abstract}

\maketitle

\section{Introduction}
In a recent paper \cite{Allison:2014zya} we discussed the possibility that new 
gauge singlet fields can have natural ultra-weak couplings amongst 
themselves 
and to Standard Model (SM) fields. 
The ultra-weak couplings, $\zeta_i$, are protected by shift symmetries
that are exact in the limit $\zeta_i\rightarrow 0$ and act
as a custodial symmetry. This ensures that
the couplings are technically natural (a similar
proposal has also appeared in \cite{Salvio:2014soa}). 
In the context of a very simple 
extension of the SM involving an ultra-weakly coupled real scalar field, we showed that a very large vacuum expectation value 
(VEV) of the scalar field can be generated.
This, in turn, induces electroweak (EW) symmetry breaking through 
the Higgs portal coupling. The large VEV of the scalar field is generated
through Coleman-Weinberg (CW) symmetry breaking \cite{Coleman:1973jx} under the assumption that the renormalized mass terms 
of the Higgs and ultra-weak field are zero.\footnote{In a field theoretic context, 
the radiative corrections to the Higgs mass that are quadratically 
dependent on the loop integral cut-off scale are not physically meaningful 
as only the renormalized $m^2$, the sum of the radiative corrections 
and the mass counter term, is measurable.}

This  ``classical scale invariance'' 
approach to the Higgs mass  is essentially empirical, 
following from the experimental 
observation of the low  mass Higgs boson. Scale invariance can be viewed as a symmetry
of a pure $SU(3)\times SU(2)\times U(1)$ SM
\cite{Bardeen:1995kv}, but it would be expected to be 
broken in the real world when including GUT, gravitational,
or any new threshold effects below the scale at which the SM couplings are defined (\eg, \cite{LlewellynSmith:1981yi,schmalz}).
However, the existence of the fundamental spin-$0$ Higgs boson makes it interesting to examine the possibility that the lower dimension
operators, \ie,  the $d=2$ renormalized 
boson mass terms (and $d=0$ cosmological constant),
are absent in the Lagrangian --- perhaps as the result of
a deeper classical scale invariance of the underlying theory.\footnote{The $d=2$ and $d=0$ terms are special in the sense that if set to zero at a high scale they remain zero in the absence of spontaneous symmetry breaking, raising the possibility that classical scale invariance is an emergent symmetry at a high scale.}
The physical Higgs mass can then be generated 
by an infrared instability involving new physics.


Whether or not the CW mechanism applies to the Higgs boson is a phenomenological question that has been explored in a large number of recent papers 
\cite{general,Hill:2014mqa}. 
However, none of the CW-Higgs  models to date have addressed 
the strong CP problem.
We consider this to be an important issue.  The usual ``invisible'' axion solution 
involves a new SM singlet scalar field $S$ that carries 
a global charge under the Peccei Quinn (PQ) symmetry 
\cite{Peccei:1977hh} and develops a very large VEV. Clearly it is 
important that the coupling of this field to the Higgs boson
does not generate an unacceptably large contribution to the Higgs mass. 
In this paper, we show that the spontaneous breaking of the PQ symmetry 
in an ultra-weak sector via  the CW mechanism can
lead to an acceptable Higgs boson mass while solving the strong CP problem.  
 
There have been two main suggestions for the nature of the PQ symmetry 
and the origin of the axion.  The DFSZ axion \cite{Dine:1981rt} 
extends the Higgs sector to include a second Higgs doublet as well 
as the complex SM singlet scalar field $S$. The Higgs doublets and 
the singlet field are charged under the PQ symmetry.  The KSVZ axion \cite{KSVZ} 
postulates that the Standard Model fields are singlets under the PQ symmetry 
and requires the addition of a ``heavy'' quark that carries non-zero 
PQ charge and couples to $S$.  In both cases the axion is identified 
with the phase of $S$ while its modulus is identified with a light 
pseudo-dilaton.\footnote{The large $S$ VEV provides the dominant source of scale breaking, 
hence the identification of the modulus of $S$ with the pseudo-dilaton.}

The origin of a light pseudo-dilaton state can be traced to the ultra-weak couplings of the $S$ field,
which are needed to avoid generating an unacceptably large mass for the Higgs 
and to enable CW breaking to generate the EW scale. Such small couplings 
are natural due to the underlying shift symmetry of $S$  
{in the limit its couplings are zero}. As a result, these couplings are
multiplicatively renormalized in the absence of gravity and 
there is no underlying expectation for their magnitude. 
Gravitational effects will generate $S$ couplings, but these
may also be small due to the shift symmetry.
Phenomenologically, the axion acquires its mass via the usual
QCD effects $m_a \sim \Lambda_{QCD}^2/ f_a$, where $f_a \equiv v_s / N_\text{DW}$ is the axion
decay constant for a domain wall number $N_\text{DW}$, 
while the dilaton acquires a mass through mixing
with the Higgs of order $m_s\sim m_h^2/v_s$.
Indeed, the observation
of the pseudo-dilaton together with the axion would provide a
smoking gun for this kind of ultra-weak mechanism.

\section{Electroweak breaking via the Coleman Weinberg  mechanism}

\subsection{The DFSZ model}

We consider the DFSZ model, which has two Higgs doublets, $H_{1,2}$, 
whose neutral components couple to the up and down 
quarks respectively and generate their masses.  We also
include the complex singlet, $S$, which carries only
the global PQ charge.
The most general classically scale invariant 
potential for $H_{1,2}$
and $S$, 
consistent with the PQ symmetry, has the form:
\bea
\label{twodub}
V(H_1,H_2,S) & = & \frac{\lambda_1}{2}|H_1|^4 +  \frac{\lambda_2}{2}|H_2|^4 + {\lambda_3}|H_1|^2|H_2|^2 
\nonumber \\
& 
+& \lambda_{4}|H_1^\dagger H_2|^2
+\zeta_{1}|S|^{2}|H_{1}|^{2}+\zeta_{2}|S|^{2}|H_{2}|^{2}\nonumber\\&+&\zeta_{3}S^{2}H_{1}^\dag H_{2}+h.c.+\zeta_{4}  \,{|S|^4},
\label{pot}
\eea
where the fields $H_{1,2}$ and $S$ are parametrized as
\begin{align}
H_1 &= \begin{pmatrix}
\phi_1^+\\
\frac{\phi_1}{\sqrt 2}e^{i\theta_1/v_1}
\end{pmatrix}, \quad 
H_2 = \begin{pmatrix}
\phi_2^+\\
\frac{\phi_2}{\sqrt 2}e^{i\theta_2/v_2}
\end{pmatrix},\nonumber\\
S &= \frac{\phi_s}{\sqrt 2}e^{i\theta_s/v_s},
\end{align}
with real moduli, $\phi_1$, $\phi_2$, and $\phi_s$,
where $\vev{\phi_1} \equiv v_1$, $\vev{\phi_2} \equiv v_2$ and $\vev{\phi_s} \equiv v_s$.
We simplify the model by taking $\lambda_4=0$;  
$\lambda_4$ will be generated by gauge interactions, 
but it remains negligibly small \cite{Hill:2014mqa}. 
We also consider the parameter range for which $v_2$ is 
small and can be treated as a perturbation, thereby 
allowing for an analytic solution to the minimisation 
conditions. A more complete study will require a numerical analysis \cite{inprep}.

There are two ways in which CW breaking can proceed. 
In the first, the dominant CW potential term is proportional 
to $\lambda_3^2$ and the interaction with the second 
Higgs field drives the quartic coefficient of the first Higgs field negative at 
some scale. This limit is equivalent to that studied in \cite{Hill:2014mqa}. 
This requires such a large $\lambda_3$ 
that there is a Landau pole in the $\sim 10$--100~TeV range. 
To avoid the appearance of a low-lying Landau pole, we therefore turn to the second possibility 
in which $\lambda_3$ is negligible and 
EW breaking is triggered by the VEV of $\phi_s$. 
The $H_1$ mass squared is then $\zeta_1 v_s^2$ and is driven negative by assuming $\zeta_1< 0$. 

To discuss this latter possibility, consider the terms 
with coefficients $\zeta_i$ in eq(\ref{pot}).
The VEV $v_s$ gives the axion decay constant $f_a = v_s / 6$ ($N_\text{DW} = 6$ in this model) and 
hence $2\times10^{9}\text{~GeV}\lesssim v_s \lesssim 10^{12}\text{~GeV}$. The singlet couplings $\zeta_{1,2,3}$ 
must therefore be very small: $\zeta_{1,2,3}\le O(m_h^2/v_s^2)$, where $m_h$ is 
the observed Higgs mass. 
For CW breaking to proceed, it is necessary 
for $\zeta_4$ to be even smaller: $\zeta_4 \le O(\zeta_{1,2,3}^2)$. 
As mentioned above, this region of parameter space is natural since the 
couplings $\zeta_i$ are forbidden in the shift symmetry limit, 
$S\rightarrow S+\delta$ \cite{Allison:2014zya}, and thus are multiplicatively renormalised. 
The stronger constraint on $\zeta_{4}$ is consistent with 
radiative corrections, as can be seen by noting the couplings 
are also forbidden by a partial scale symmetry $S\rightarrow\lambda S$, 
where $\zeta_{1,2,3}$ scale as $\lambda$ while $\zeta_{4}$ scales 
as $\lambda^{2}$. If the symmetry is broken (perhaps by gravity) 
by a term scaling as $\lambda$, the relative ordering of $\zeta_{4}$ results.
 
Even though the $S$ couplings are all extremely small, 
CW breaking in the $S$ sector is still possible. 
It is convenient to consider the phenomenologically relevant 
limit in which the term proportional to $\zeta_{2}$ provides the dominant CW term.  
It is in this limit that the additional Higgs states coming from the second 
Higgs doublet are heavy enough to have escaped detection to date~\cite{2HDM}.\footnote{We treat the effect of the $\zeta_3$ term perturbatively 
as it drives the $H_{2}$ VEV, which we have assumed to be in the p
erturbative regime.} In this limit the potential, 
including the dominant one-loop correction, can be written as:
\begin{align}
V(\phi_1,\phi_s) 
&\approx \frac{\lambda_1}{8}(\phi_1^{2}+\alpha\phi_s^{2})^{2}\nonumber\\
&+ \frac{1}{64\pi ^2}\left( \zeta_{2} \phi_s^2\right)^2\left[\ln \left( 
\frac{\zeta_2\phi_s^2 }{\bar M^2} \right)-\frac{3}{2}\right],
\label{pot1}
\end{align}
where $\alpha=\zeta_{1}/\lambda_{1}$, 
$\bar M^{2}=2M^{2}e^{-16\pi^2 (\zeta_{4}-\lambda_{1}\alpha^{2}/2)/\zeta_{2}^{2}}$, 
and $M$ is the scale at which the couplings are defined. 
This has a minimum at (minimization of a similar single
Higgs potential is discussed in
\cite{Allison:2014zya}):
\begin{align}
v_{s}^{2}&=\frac{e\bar M^{2}}{\zeta_2},\qquad
v_{1}^{2}=-\alpha v_{s}^{2}.
 \label{vevs}
\end{align}
Finally, $v_2$ is driven by the term proportional 
to $\zeta_{3}$ in eq(\ref{pot}), giving:
\beq
v_{2}\approx-\frac{\zeta_{3}}{\zeta_{2}}v_{1}.
\eeq

In the region of parameter space considered here, 
mixing between states is small and the observed Higgs $h$ 
is approximately $\phi_1$. Similarly, the other neutral Higgs $H$ 
and the pseudo-dilaton are approximately 
$\phi_2$ and $\phi_s$, respectively. 
Then a straightforward calculation gives:
\begin{align}
m_{h}^{2}&\approx -\zeta_1 v_s^2 \approx \lambda_{1}v_1^2,\nonumber\\
m_H^{2}&\approx \zeta_2 v_s^2 / 2,\nonumber\\
m_s^{2}&\approx\zeta_{2}^{2}v_s^{2}/8\pi^{2}.
\label{masses1}
\end{align}

Determining the charged Higgs masses is 
more subtle as they only acquire mass 
via the term proportional to $\zeta_{3}$ 
in eq(\ref{pot}). This happens because all the other 
terms are functions of $|H_{1}|^{2}$ and $|H_{2}|^{2}$, so:
\beq
\partial^{2}V/\partial \phi_{1,2}^{+}\partial \phi_{1,2}^{-}\propto \partial V/\partial \phi_{1,2}=0.
\eeq
As a result, we find the ``uneaten'' 
charged Higgs state has a mass:
\beq
m_{H^{\pm}}^{2}\approx -(v_{1}/v_{2})(\zeta_{3}/2)v_s^{2}=\zeta_{2}v_s^{2}/2.
\eeq
Finally we turn to the phases of the fields. One combination,
\beq
\theta_Z \propto \theta_1 v_1+\theta_2v_{2},
\eeq
provides the longitudinal component of the $Z$ boson. 
An orthogonal combination given by:
\beq
\theta_A=( -\theta_1/v_1 + \theta_2/v_2 + 2\theta_s/v_s)/N,
\eeq
where:
\beq
N^{2}=1/v_{1}^{2}+1/v_{2}^{2}+4/v_{s}^{2},
\eeq
gets mass from the $\zeta_{3}$ term. Its mass is given by:
\beq
m_A^{2}=-(\zeta_{3}/2)v_{s}^{2}v_{1}v_{2}N^{2}\approx- 
(\zeta_{3}/2)v_{s}^{2}v_{1}/v_{2}=\zeta_{2}v_{s}^{2}/2.
\eeq
The orthogonal state to $\theta_Z$ and $\theta_A$ 
is the axion. The axion only gets its mass from QCD effects, as usual. 
 
To summarise, we have ($\zeta_1 < 0$):
\begin{align}
m_H^2&=m_{H^{\pm}}^{2}=m_A^{2}=-\frac{\zeta_{2}}{2\zeta_{1}}m_{h}^{2},\nonumber\\
m_s^2&=\frac{\zeta_{2}}{4\pi^{2}}m_H^{2}=-\frac{\zeta_{2}^{2}}{8\pi^{2}\zeta_{1}}m_{h}^{2}.
 \label{masses}
\end{align}
 
\subsection{The KSVZ  model}

In the KSVZ model, the SM states are PQ singlets. However, 
the SM singlet field $S$ interacts with some new heavy quark 
$X_{L,R}$, which is vector-like with respect to the SM gauge 
group but carry PQ charge, via the Yukawa interaction
\beq
L_{KSVZ}=-f\bar X_{L}SX_{R}-f^{*}\bar X_{R}S^{\dagger} X_{L}.
\label{KSVZ}
\eeq
Imposing classical scale invariance, the scalar potential has the relatively simple form
\be
V(H,S)=\frac{\lambda}{2}|H|^{4}+\eta_{1} |S|^{2}|H|^{2}+\eta_{2}|S|^{4},
\label{pot2}
\ee
where $H$ is the SM Higgs doublet.

Following from the non-observation of additional coloured 
states up to the TeV range and the need to keep the Higgs light, 
one sees from eqs.(\ref{KSVZ}) and (\ref{pot2}) that the largest 
coupling to the $S$ field is $f$ and the associated dominant 
loop correction to the $S$ potential involves the new heavy quark. 
As a result, the loop correction contributes to the potential 
with a relative minus sign compared to that of eq(\ref{pot1}) in the DFSZ case.  
This does {\it not} give rise to one-loop EW breaking because, 
if it triggers EW breaking, it drives the Higgs VEV to an unacceptably large scale.
 Avoiding this problem requires an additional CW radiative correction 
 with the opposite sign to dominate. Such a term could arise if there 
 are additional SM singlet fields. 
 It could also possibly be engineered at the two-loop level by
fermion loops, similar to a model discussed in \cite{Hill:2014mqa}. 
We do not explore these possibilities further here.

\section{Phenomenological implications}

The DFSZ model requires the 
extension of the SM spectrum to include a second doublet 
of Higgs fields and a complex singlet $S$ which contains 
the axion $a$ and the pseudo-dilaton $\phi_s$. The ultra-weak 
couplings $\zeta_i$ ensure that for collider experiments 
the phenomenology of the model is just that of the Type II 
two Higgs doublet model (2HDM) with the common mass scale 
of the additional Higgs states $\bparen{H,A,H^\pm}$ determined 
by the ratio $R \equiv m_H/m_h \simeq \sqrt{\zeta_2/|2\zeta_1|}$. 
In the 2HDM, additional Higgs states with masses of roughly $350$~GeV or above, 
which corresponds to $R \gtrsim 3$, are allowed in significant 
regions of parameter space~\cite{2HDM}.\footnote{Note that the convention in studying the Type II 2HDM is to have $H_2$ couple to the up-type quarks rather than $H_1$. Therefore when applying 2HDM limits to this model, one should use the definition $\tan \beta \equiv v_1 / v_2$ rather than the usual $\tan \beta \equiv v_2 / v_1$.} 

At the same time, an approximate upper bound $R \lesssim 5$ 
comes from the requirement that one does not re-introduce the 
little hierarchy problem due to the coupling between the light 
and heavy Higgs sectors. This requires the masses of the heavy 
Higgs states to be smaller than $O(600\text{~GeV})$.

In the usual implementation of the DFSZ model, 
the pseudo-dilaton $\phi_s$ is very heavy with a mass of $\mathcal{O}(v_{s})$. 
The novel feature of the model discussed here is that $\phi_s$ is very light. 
From eqs(\ref{vevs}) and (\ref{masses1}), we have:
\beq
\zeta_{1}=-\frac{m_h^2}{v_s^2} \approx -1.6 \times 10^{-20}\parenfrac{10^{12}\text{~GeV}}{v_s}^2,
\eeq
and hence from eqs.(\ref{vevs}) and (\ref{masses}):
\begin{align}
m_s^2&=-\frac{\zeta_1}{2\pi^2}\parenfrac{m_H}{m_h}^4 m_h^2\nonumber\\
&\simeq 32 \parenfrac{10^{12}\text{~GeV}}{v_s} \parenfrac{R}{3}^2 \text{~eV}.
\label{dfsz}
\end{align}
Since the pseudo-dilaton is light and couples to 
quarks through its mixing with the SM Higgs, 
one way to detect it is through fifth force experiments. 
However, using the estimate for the coupling strength of the pseudo-dilaton to protons:
\begin{align}
\alpha_5 \sim \frac{1}{2\pi}\sparen{\parenfrac{2 m_u + m_d}{v_s} }^2
\end{align}
and $\sim 1/m_s$ for the effective range of the dilaton exchange force, 
it can be seen that the pseudo-dilaton lies outside 
the region excluded by Casimir-force and neutron scattering 
experiments~\cite{Salumbides:2013dua}.

The axion couples to electromagnetic fields through the axial anomaly
in the usual way, $\sim c(a/v_s)(\alpha/4\pi)F_{\mu\nu}\widetilde{F}^{\mu\nu}$.  Likewise,
the dilaton couples as $\sim c'(\phi_s/v_s)(\alpha/4\pi)F_{\mu\nu}{F}^{\mu\nu}$,
with $c,c' \sim {\cal{O}}(1)$.
A detailed analysis of the detectibility and limits from the electromagnetic
coupling for the dilaton goes beyond the scope of this paper.
It is possible that future  terrestrial ``5th-force'', nuclear and RF-cavity 
experiments can be devised to look for the pseudo-dilaton 
directly, but this remains unexplored.
It is possible that future 
terrestrial experiments can be devised to look for the pseudo-dilaton 
directly, but this remains unexplored. At present, the only way to 
constrain the pseudo-dilaton is through its cosmological influences, 
which we turn to a discussion of now.

\section{Cosmology of the pseudo-dilaton}

If the $S$ field acquires its VEV before inflation, 
the energy density of the pseudo-dilaton will be diluted away. 
This is the case if the dilaton mass is larger than the Hubble 
parameter during inflation, which requires a low scale of inflation:
\begin{align}
{V_\text{inf}^{1/4}} \lesssim 10^{5}\parenfrac{10^{12}\text{~GeV}}{v_s}^{1/2} R \text{~GeV},
\label{inflationscale}
\end{align}
and if the reheat temperature is sufficiently low such that
 the PQ symmetry is not restored after inflation.
On the other hand, if the PQ symmetry breaking occurs 
after inflation, there will be energy stored in the dilaton 
potential that will be released after inflation 
(the Polonyi problem \cite{Coughlan:1983ci}) 
in the form of dilaton oscillations. We shall consider 
both cases in turn, starting with the latter case.

\subsection{High scale inflation}

The energy stored in the dilaton potential depends on the 
initial value (VEV) of the dilaton. For the case that the Hubble 
parameter during inflation is much larger than the dilaton mass, 
the dilaton will perform a random walk of step length 
$H_\text{inf}/2\pi$ in each Hubble time. The maximum dilaton 
energy corresponds to the largest initial value of $\vev{\phi_s}$, 
which in turn corresponds to the case of the maximum Hubble 
parameter during inflation, $H_\text{inf} \sim 10^{14}$~GeV, 
consistent with the BICEP2 result \cite{Ade:2014xna}. 
To be conservative, let us consider this extreme case since all 
others will have a smaller amount of energy stored in the dilaton 
and will be more weakly constrained. 
For 70 e-folds of inflation, one may expect the initial value 
of the dilaton field to be given by $\vev{\phi_{s}}_i \sim 10^{14}$~GeV.

After inflation and reheat, $\vev{\phi_s}$ begins to oscillate 
when its effective mass becomes larger than the Hubble parameter.  
In the presence of a thermal bath, $\phi_s$ obtains a large 
thermal mass~\cite{Mukaida:2013xxa}
\begin{align}
\label{eq:thermalmass}
m_{s\text{,th}}^2 \simeq \frac{\zeta_2}{6} T^2,
\end{align}
where we have neglected all but the largest 
coupling of $\phi_s$ to thermalized particles.  
For a sufficiently high reheat temperature, 
the thermal mass eq.\eqref{eq:thermalmass} dominates 
the dilaton potential and the dilaton oscillates 
about a zero VEV when it begins to roll.  
The roll begins when $m_{s,\text{th}} \sim H$, 
corresponding to the temperature
\begin{align}
T_\text{roll}\approx 5\times 10^7 R \parenfrac{10^{12}\text{~GeV}}{v_s}\text{GeV}.
\end{align}

As the universe expands, the energy 
density in the dilaton at the beginning of the roll,
\begin{align}
\rho_{s,\text{roll}} \simeq \frac{\zeta_2 T_\text{roll}^2}{12} \vev{\phi_s}_i^2,
\end{align}
redshifts as radiation, \ie\ $\propto T^4$~\cite{Mukaida:2013xxa}. 
This is faster than the matter redshift that one might expect
 because the temperature-dependent thermal mass also redshifts.  
 As the temperature drops down to $T \sim 20R$~GeV, the thermal 
 mass of the dilaton becomes comparable in size to the 1-loop 
 term in the potential~eq.\eqref{pot1} and the minimum at $\vev{\phi_s} = v_s$ appears.  
 However, the tunnelling rate to the true vacuum is low and the 
 dilaton continues to oscillate about a zero VEV.\footnote{The amplitude of the oscillations scales $\propto T$ from its initial 
 value $\vev{\phi_s}_\text{i}$ at $T_\text{roll}$. At $T \sim 20R$~GeV, the 
 amplitude of the oscillations is also too small to reach the minimum at $v_s$.}

The dilaton oscillates about a zero VEV until the 
EW symmetry is ultimately broken at the 
temperature when QCD becomes non-perturbative and 
drives the quark condensate, which in turn gives 
masses to the W and Z bosons as well as the Higgs. 
Once the temperature drops below the masses of these bosons, 
the stabilizing thermal mass term for the dilaton 
rapidly vanishes due to the Boltzmann 
suppression~\cite{Mukaida:2013xxa} and the 
dilaton rolls to its true minimum at $v_s$.

After the temperature has dropped below about $T \sim 10R$~GeV 
and until the EW symmetry is broken, the energy 
density of the universe is dominated by the potential energy 
in the Higgs and dilaton fields.  This will give rise to a period 
of thermal inflation with roughly $\ln(10R\text{~GeV}/200\text{~MeV})\sim 5$ 
e-folds of inflation. 
This period of thermal inflation does not 
affect the density perturbations coming from the 
initial stage of slow-roll inflation and still 
allows for successful baryogenesis.






Once the EW symmetry is broken, 
the potential energy in the Higgs field 
reheats the thermal plasma. Since the Higgs' 
couplings to the plasma are $\mathcal{O}(1)$, 
the reheating is efficient and gives a reheat 
temperature of {$T_\text{reh} \sim 10R$~GeV.}  
Meanwhile, the potential energy in the dilaton,
\begin{equation}
\label{eq:storedE}
\rho_{s} \simeq \frac{\zeta_2^2 v_s^4}{{256}\pi^2} {\simeq \frac{R^4 m_h^4}{64\pi^2}},
\end{equation}
is released as a coherent oscillation of the field 
that redshifts as matter, \ie\ $\propto T^3$.

This energy density is large enough that it will quickly
dominate the energy density of the universe, thereby 
ruining late-time cosmology, unless it is somehow dissipated.  
{This indeed happens because of a resonant enhancement 
of the scattering rate of the coherent state of zero momentum oscillating 
dilatons on the thermal background.}

{To illustrate this consider the process $s+c \to c \to \text{SM states}$ 
involving the scattering of the dilaton off the distribution 
of charm quarks. Since the dilaton mass is so small, 
the intermediate $c$ is nearly on-shell and its propagator 
is dominated by its thermal width $\Gamma_c \simeq G_F^2 m_c^5/(192\pi^3)$.\footnote{Here we neglect the finite temperature corrections and use the 
zero temperature width. This is valid for temperatures $T \lesssim m_c$, 
at which the dissipation rate is sufficiently large anyway.} 
Since this width is small, there is an enhancement of the scattering 
rate that leads to a thermal dissipation rate of the dilaton given by~\cite{Mukaida:2012qn}
\begin{equation}
\Gamma_s \simeq \frac{\sqrt 2m_c^4}{\pi^{3/2} v_s^2 \Gamma_c} \parenfrac{T}{m_c}^{1/2} e^{-m_c/T}.
\label{dissipation}
\end{equation}
This rate exceeds the Hubble expansion rate over some range 
of temperatures $T_* < T \lesssim m_c$ for $v_s \lesssim 5 \times 10^{14}$~GeV. 
Thus the dilaton oscillations are dissipated for all $v_s$ of interest.}

{Note that the inverse dilaton production processes, such as $g + q \to q \to q + s$ 
where $g$ is a gluon and $q$ is a thermalized quark, 
do not have a resonant enhancement because none of 
the reactants are zero momentum coherent states. 
Due to the low reheat temperature $T_\text{reh} \sim 10R$~GeV, 
the number density of the top quark is exponentially suppressed 
and it is the bottom quark scattering that gives the largest 
rate of dilaton production. An estimate of this rate for $T \gtrsim m_b$ is:
\begin{align}
\Gamma_s^\text{prod} \simeq \frac{9 \zeta(3)}{\pi^2 } \parenfrac{m_b}{v_s}^2 \alpha_s T,
\end{align}
which produces a dilaton population:
\begin{align}
\frac{n_s}{n_s^\text{eq}} \sim \left. \frac{\Gamma_s^\text{prod}}{H} \right|_{T = m_b} \sim 0.4 \parenfrac{2 \times 10^9\text{~GeV}}{v_s}^2.
\end{align}
If the dilaton is sufficiently long lived, 
it is non-relativistic today with an abundance:
\begin{equation}
\label{eq:omegasmass}
\Omega_s \sim 0.3 \parenfrac{R}{3}^2 \parenfrac{7 \times 10^9\text{~GeV}}{v_s}^3.
\end{equation}
To constitute dark matter, however, the dilaton must be 
stable on cosmological timescales. The dominant direct 
decay mode of the dilaton is to two axions with the decay rate:
\begin{align}
\label{eq:axiondecay}
\Gamma_{s\rightarrow aa} = \frac{1}{32\pi}\frac{m_s^3}{v_s^2} \simeq \frac{R^6 m_h^6}{64\sqrt 2 \pi^4 v_s^5},
\end{align}
giving a lifetime:
\begin{align}
\label{eq:dilatonlifetime}
\tau_s \simeq 3.4 \times 10^{18} \parenfrac{3}{R}^6 \parenfrac{v_s}{7 \times 10^{10}\text{~GeV}}^5\text{sec}.
\end{align}
Constraints on decaying dark matter require 
the lifetime to be on the order of 100~Gyr ($3 \times 10^{18}$~sec) 
or longer~\cite{DMdecay}. Thus for $v_s \sim 7 \times 10^9$~GeV 
for which a significant dilaton population is produced, the dilaton 
is unstable on cosmological time scales and cannot be dark matter. 
Conversely, for $v_s \gtrsim 7 \times 10^{10}$~GeV for which the 
dilaton is sufficiently stable, dilaton production is negligible.}

{The axion, however, provides a very plausible cold dark matter candidate. 
The energy density in the coherent oscillations (zero mode) of the 
axion through vacuum realignment is~\cite{axionDM}:
\begin{align}
\label{eq:axionDM}
\Omega_a h^2 \simeq 0.236\theta_i^2 f(\theta_i) \parenfrac{v_s / N_\text{DW}}{10^{12}~\text{GeV}}^{7/6},
\end{align}
where $N_\text{DW} = 6$ for this model, $\theta_i$ 
is the initial misalignment angle, and 
the function $f(\theta_i) = \sparen{\ln( e / (1-\theta_i^2/\pi^2))}^{7/6}$ 
encodes the anharmonic effect.
Meanwhile, the higher momentum axion modes and the axions 
produced in the decay of strings and domain walls contribute 
a comparable amount to the energy density as vacuum 
realignment~\cite{Sikivie:2009zz}. For $v_s \sim 10^{12}$~GeV, 
the axion can therefore provide all of the dark matter.}

There remains the question of how an unacceptable 
energy density from the domain walls produced after 
the PQ breaking transition can be avoided. 
Since $N_{DW}=6$, the energy density in stable 
domain walls is many orders greater than the critical 
energy density for closing the universe and completely 
unacceptable. However, small PQ breaking can cause the 
walls to decay and hence avoid the problem while 
preserving the axion solution to the strong CP problem. 
To see how this can happen, we note that the most general 
potential $V(H_{1},H_{2},S)$ includes the terms 
$\lambda_{5}(H_{1}^{\dagger} H_{2})^{2}+
{\zeta_{5}}S^{2}H_{2}^{\dagger} H_{1}+
{\zeta_{6}}S^{4}+h.c$ 
that break the PQ symmetry and splits the degeneracy 
of the $Z(N_\text{DW})$ discrete symmetry that leads to the domain wall problem. 
Note that these couplings multiplicatively renormalise and so, 
following the discussion above, we conclude they can naturally 
be arbitrarily small. 
{Indeed, with PQ breaking of a similar 
magnitude to scale breaking ($\lambda_{5}\sim \zeta_{1,2,3}$) 
these terms are in the range needed to solve the domain 
wall problem without disturbing the axion solution of the strong CP problem~\cite{Sikivie}.} 

{In summary, the large thermal mass of the dilaton 
produces a period of thermal inflation with 
approximately 5 e-folds after the usual slow roll inflation.  
For all $v_s$ of interest, the interactions of the light dilaton 
with the thermal bath dissipate the energy in its coherent oscillations. 
A significant relativistic population of dilatons is produced in the 
region of parameter space with $v_s \lesssim 7 \times 10^9$~GeV, 
but the dilaton is too short-lived in this region to be dark matter; 
the dilaton can therefore only have a negligible contribution to 
dark matter. The axion, however, can comprise all of the dark matter for $v_s \sim 10^{12}$~GeV.}

\subsection{Low scale inflation}

{In the case that eq.\eqref{inflationscale} 
is satisfied and the reheat temperature is 
sufficiently low ($T_\text{reh} \lesssim 100$~GeV) 
that the dilaton does not obtain a thermal mass that 
forces it to roll to $v_s = 0$, the PQ symmetry remains 
broken during and after slow roll inflation. As a result, 
the energy density in the dilaton oscillations are driven 
exponentially small and the axion field gets homogenized 
by the expansion of the universe during the inflationary 
phase, thereby preventing the formation of domain walls~\cite{Sikivie:2009zz}. 
The usual result eq.\eqref{eq:axionDM} for the axion contribution 
to the energy density from vacuum realignment still holds\footnote{The contribution from the quantum fluctuations of the axion field during inflation, which are included by making the replacement $\theta_i^2 \to \theta_i^2 + \sigma_\theta^2$ in eq.\eqref{eq:axionDM} where $\sigma_\theta^2 \simeq (H_I/2\pi f_a)^2$~\cite{axionDM}, are negligible for $H_I$ satisfying eq.\eqref{inflationscale}.} and 
the axion can provide all of the dark matter for 
$v_s \sim 10^{12}$~GeV and $\theta_i \sim 1$.\footnote{As discussed in~\cite{axionDM}, with moderate 
fine-tuning to give $\theta_i \simeq \pi$, the axion 
can provide all of the dark matter for smaller values 
$v_s$ due to the anharmonic effect.} 
The dilaton, however, never plays a significant role in cosmology.}

\section{An ultra-weak DFSZ SUSY model}
{Classical scale invariance of the low-energy theory does not apply if there 
are heavy states coupled to the Higgs, such as Grand 
Unified states, with mass below the scale at which the SM couplings are defined. 
These introduce radiative contributions to the 
Higgs mass that are proportional to the mass of 
the heavy GUT states. Unlike the radiative 
corrections to mass simply proportional to 
the cut-off scale, these GUT corrections involve 
a logarithmic dependence on the scale at which 
they are measured and thus are physical.  To avoid the hierarchy problem, 
the model discussed above must therefore 
not have a stage of Grand Unification.}

{It {\it is} possible to include a stage of Grand Unification 
in a scale invariant theory by super-symmetrizing the model 
so that the contribution to the Higgs mass coming from interactions 
with the heavy GUT states, although present, are acceptably small. 
 As we sketch below, CW breaking in the ultra-weak sector associated 
 with the axion can readily be extended to a supersymmetric theory.}

The states of the DFSZ model neatly correspond to 
the non-SUSY states of the ($N$=1) NMSSM, so a 
supersymmetric version of the model can be constructed easily. 
After imposing the PQ symmetry, the allowed couplings are 
more restricted than those of the NMSSM and correspond to 
those recently discussed in \cite{Dreiner:2014eda}. The only 
term in the superpotential $W$ involving the  $S$ field is: 
\beq
W=\zeta_{1 }\hat S\hat H_{1}\hat H_{2},
\label{W}
\eeq
where the scalar components of the super fields 
$\hat S$ and $\hat H_{1,2}$ are the $S$ field and the Higgs doublets. 

Due to the constraints of supersymmetry, the 
model is classically scale invariant in the 
absence of SUSY breaking. We are interested in 
the case that $\zeta_{1}$ is ultra-weak, which is 
natural due to the underlying shift symmetry 
when $\zeta_{1}$ is zero.
Allowing for SUSY breaking, the only other 
terms involving just these fields are the 
soft terms:\footnote{{Soft terms can be generated in 
a classically scale invariant theory through 
spontaneous breaking, for example via gaugino condensation.}}
\beq
V(H_{1,2},S)=m_s^{2}|S|^{2}+m_1^{2}|H_{1}^{2}|+m_2^{2}|H_{2}^{2}|+T_{cl}S H_{1}H_{2}.
\label{V}
\eeq

The quartic scalar terms coming from eq.(\ref{W}) 
are positive semi definite, so the only possibility 
for dynamical SUSY breaking is through the soft terms. 
Including radiative corrections, $m_{s}^{2}$ can be 
driven negative by radiative corrections proportional
 to $\zeta_{1}^{2}m_{{1,2}}^{2}$. This triggers $v_s$ 
 at a scale close to the point at which the mass is zero, which can be very large, as required. However, this requires that the starting value of $m_s^{2}$ should be ultra-small relative to $m_{{1,2}}^{2}$. An ultra-small mass is natural if there is an underlying shift symmetry, which can readily happen if, for example, SUSY is broken in a hidden sector and SUSY breaking is communicated to the $S$ field by gravitational effects while the SM states receive their SUSY breaking masses via gauge mediation. In this case, the soft $S$ mass and the graviton will be much lighter than the SUSY breaking masses $m_{1,2}^{2}$ in the visible sector.
The dimensional transmutation mechanism in the UW sector 
provides an economical and elegant origin for the axion 
decay constant that does not require the inclusion of 
an O'Raifeartaigh term involving an explicit mass scale \cite{Dreiner:2014eda}.

The SUSY phenomenology of the model is essentially that 
of the minimal supersymmetric SM, theMSSM (with gauge mediation) 
because the additional couplings of the Higgs to the 
singlet sector are ultra-weak and hence insignificant, 
apart from providing the origin of the $\mu$ term of the 
MSSM, $\mu=\zeta_{1}v_{s}$, c.f.\ eq(\ref{W}). In this case, 
EW breaking proceeds in the usual way through radiative 
corrections that, due to the top Yukawa coupling, drive 
the soft Higgs mass squared negative \cite{Ibanez:1982fr, Inoue:1982pi,AlvarezGaume:1983gj}.

The LSP is the axino, the fermion component of $S$, with a mass:
\beq
m_{\hat S}=\frac{\mu v_{1}v_{2} }{v_s^{2}}\sim 10^{-9}\left(\frac{10^{12}\text{~GeV}}{v_{s}}\right )^{2}\text{eV}
\eeq
generated by the see-saw mechanism through 
its coupling to the Higgsinos. The decay of 
the lightest MSSM SUSY state to the gravitino 
or axino is so slow that it does not occur within 
the detector and does not change the MSSM phenomenology. 
The dark matter component that ends up in the axino 
depends on the MSSM parameter choice and has been 
discussed extensively elsewhere.  

Due to the quartic couplings associated with the 
superpotential term in eq(\ref{W}), the Higgs obtain 
$S$ dependent masses as in the non-supersymmetric 
DFSZ model. As a result, Higgs oscillations are driven 
by the dilaton oscillations in the manner discussed above. 
The energy in the dilaton fields is converted to energy 
in the SM sector at a time before nucleosynthesis and 
does not significantly change the usual MSSM cosmology. 
\vspace{0.5 cm}

\section{Summary and conclusions}

{The discovery of a Higgs scalar with properties 
very close to that predicted by the SM, 
together with the absence of any indication for 
physics beyond the SM, has led to a 
re-evaluation of the need for such physics to 
solve the hierarchy problem. Formally, as a pure field theory, 
the SM has no hierarchy problem because the radiative 
corrections to the Higgs mass squared that are 
quadratically dependent on the cut-off are not 
physical; only the renormalised mass is measurable, 
so any value of $m$ is possible and only the empirical choice $m=0$, 
which corresponds to classical scale invariance of the theory, is special.
}

{With this motivation, we discussed how the SM could 
result from a classically scale invariant theory 
that also addresses the major questions left 
unanswered by the SM. While there has been extensive 
discussion of the possible origin of baryogenesis, 
dark matter and inflation, very little attention has been
paid to the the strong CP problem in such 
theories. In this paper, we showed how a scale invariant 
version of the DFSZ model can spontaneously generate 
the large PQ scale through an ultra-weakly coupled 
sector involving a complex SM singlet scalar field $S$. 
As discussed in  \cite{Allison:2014zya}, such an ultra-weak 
sector involving gauge single fields is technically 
natural due to an underlying approximate shift or scale symmetry.}

{Due to the ultra-weak couplings, the DFSZ extension of 
the SM contains an anomalously light pseudo-dilaton as 
well as the usual axion, which come from the complex 
scalar $S$. Despite the ultra-weak couplings, there is 
no Polonyi problem associated with the $S$ field due to a 
resonant enhancement of the scattering of the coherent $S$ 
state off the thermal background after the PQ and 
EW phase transition are triggered. 
Unusually, the PQ phase transition occurs at the EW 
scale in this model. Meanwhile, the dilaton 
production cross section does not have the 
resonant enhancement and its abundance is typically 
negligible. Dark matter can be in the form of axions 
produced via a combination of vacuum alignment and 
decay of axion domain walls. Such decay is possible 
due to additional PQ breaking terms, which can be 
consistent with the axion solution to the strong 
CP problem as long as they are also ultra weak and 
have a strength comparable to the scale breaking terms.}

{Due again to the ultra-weak couplings of the singlet fields, 
the phenomenology of the model is that of the usual two-Higgs 
doublet extension of the SM. The most significant constraint 
on the additional heavy Higgs states comes from the requirement 
that the little hierarchy problem is not re-introduced. 
It may be possible to search for the ultra-light dilaton 
along the lines suggested in \cite{Buchmuller:1989rb}, but 
this remains to be studied.}

{Finally, we outlined the construction of a scale-invariant 
SUSY version of the model that can accommodate a stage of 
Grand Unification without re-introducing the hierarchy problem.
 It provides a simple origin for the $\mu$ term and the LSP is the axino, 
 the fermion component of the super field that contains the DFSZ complex 
 scalar field $S$. However, since the decay of the lightest 
 MSSM state to the LSP is extremely slow, the collider 
 phenomenology of the model is just that of the MSSM with 
 gauge mediated SUSY breaking.}
 
\vskip 3pt
{\bf Acknowledgements} One of us (GGR) would like to thank I.Antoniadis, W.Buchmuller, 
S. Davidson and L. Alvarez-Gaume for useful comments. 
Part of this work was done at Fermilab, operated by Fermi Research Alliance, 
LLC under Contract No. DE-AC02-07CH11359 with the United States Department of Energy. GGR would like to thank the Leverhulme foundation and Fermilab for support during the course of this research.

\end{document}